\documentclass[aps,amsmath,amssymb,amsfonts,prd,nofootinbib,reprint,nobibnotes]{revtex4-2}
\pdfoutput=1 

\usepackage{graphicx}
\usepackage{hyperref}

\newcommand{\comment}[1]{}
\newcommand{\beq}[1]{\begin{equation}\label{#1}}
\newcommand{\eeq}{\end{equation}}

\newcommand{\C}{\mathcal{C}}

\newcommand{\del}{\partial}

\newcommand{\M}{\mathcal{M}}
\newcommand{\bea}{\begin{eqnarray}}
\newcommand{\eea}{\end{eqnarray}}
\newcommand{\N}{\mathcal{N}}
\newcommand{\w}{\wedge}
\renewcommand{\t}{\tilde}

\begin{document}

\title{Holographic Complexity in String and M Theory}

\author{Andrew R.~Frey}
\email{a.frey@uwinnipeg.ca}
\affiliation{Department of Physics and Winnipeg Institute for
Theoretical Physics, University of Winnipeg\\
515 Portage Avenue, Winnipeg, Manitoba R3B 2E9, Canada}

\begin{abstract}
Thus far, the literature regarding holographic complexity 
almost entirely focuses 
on the context of $(d+1)$-dimensional anti-de Sitter
spacetime rather than the full higher-dimensional gauge/gravity duality
in string or M theory. We provide a framework to study 
holographic complexity in the full duality, explaining the relation of
complexity functionals in the higher-dimensional theory to those in 
anti-de Sitter spacetime and when complexity functionals can apply universally
to gauge/gravity dualities rather than a specific dual pair. We also show
that gauge invariance constrains boundary terms for complexity functionals,
using the ten- and eleven-dimensional supergravity actions as key examples.
Finally, we propose new universal complexity functionals following these
considerations, including a revised gauge-invariant action complexity.
\end{abstract}

\maketitle

\section{Introduction}

Quantum information theory has taken an important role in the study of 
quantum gravity, as it has for many subjects in physics. After the work
of \cite{hep-th/0603001,hep-th/0605073} on entanglement entropy, there is
now a large literature on quantum information theory in holography, physical
systems in which gauge theories are equivalent to gravitational ones,
particularly the anti-de Sitter spacetime/conformal field theory
correspondence \cite{hep-th/9711200}. 

Circuit complexity (or computational complexity) is a quantum information
theoretic measure of the geodesic distance in Hilbert space of 
the state of a system 
from a given reference state. Holographic complexity is a gravitational 
functional that corresponds to the complexity of the dual gauge theory; 
Susskind and co-workers 
\cite{arXiv:1403.5695,arXiv:1406.2678,arXiv:1509.07876,arXiv:1512.04993}
realized that complexity plays an analogous role to entropy
in quantum gravity (for a pedagogical explanation, see \cite{arXiv:1810.11563}).
The most studied holographic complexity functionals are based on volume 
(either of a maximal spacelike slice or a specific region of spacetime
\cite{arXiv:1610.02038}) and the action, although more recent
developments \cite{arXiv:1901.00014,arXiv:2111.02429,arXiv:2210.09647}
indicate that an infinite family of functionals is suitable to describe
holographic complexity, in line with the fact that there is an infinite 
set of distance measures on Hilbert space. For a review of holographic
complexity and many references, see \cite{arXiv:2110.14672}.

So far, studies of holographic complexity have almost exclusively focused on
asymptotically anti-de Sitter (AdS) spacetimes 
(with notable exceptions of 
\cite{arXiv:2109.06883,arXiv:2111.14897,arXiv:2311.18804,arXiv:2401.04158}).
However, this approach is intrinsically limited in answering questions 
related to the embedding of the AdS spacetime in 
string/M theory, an important aspect of the full duality needed to determine
features of the field theory such as the detailed spectrum and $R$ symmetry. 
For example, a purely geometric functional in AdS spacetime cannot
distinguish the conformal field theories (CFTs) 
dual to AdS$_5\times S^5$ and AdS$_5\times T^{1,1}$,
despite the difference in field content,
while a geometric functional in the full 10D may. Furthermore, many
nonconformal gauge/gravity duals have explicitly higher-dimensional structure,
such as dualities involving brane polarization (including the $\N=1^*$ gauge
theory). More fundamentally, gauge/gravity dualities offer a nonperturbative
definition of higher-dimensional (super)gravity and string/M theory, so a
higher-dimensional formulation of holographic complexity is an opportunity
to improve our understanding of those higher-dimensional gravities.

\comment{
Nonetheless, there are many reasons to consider holographic complexity in
higher-dimensional spacetimes, which may contain AdS factors (overall or
in an asymptotic region) or serve as the gravitational side of a more general
gauge/gravity duality. }

This note collects several distinct but related observations as 
prerequisites to a systematic study of holographic complexity in the full 
context of gauge/gravity duality in string/M theory. First, we discuss
a basic framework for holographic complexity in different dimensions
and two classes of complexity functionals. Then we show that gauge invariance
for complexity functionals with Chern-Simons terms requires specific
contributions on the boundary of the region where complexity is evaluated.
This is particularly important for the action complexity in
10D and 11D supergravity (SUGRA).
Finally, we propose complexity functionals in 10D and 11D supergravity
which generalize the typical volume (known as CV or CV2.0) 
and action (known as CA) complexities
as used in AdS spacetimes. We conclude by discussing important questions 
and potential applications.


\section{Complexity across dimensions}

We consider general complexity functionals as described by the 
``complexity=anything'' program \cite{arXiv:2111.02429,arXiv:2210.09647}.
Namely, holographic complexity is a functional evaluated on a
codimension-0 region $\M$ determined by optimization of another functional;
the past and future boundaries of $\M$ intersect on a slice $\Sigma$ of the
holographic boundary (defined as a cutoff surface at large radius)
where the complexity is to be evaluated.
Complexity functionals evaluated on a codimension-1 surface are given by 
a special case of the optimization procedure. Such functionals 
in AdS spacetimes that depend
only on the metric and its derivatives satisfy two key properties of
the complexity\footnote{Linear growth at late times in AdS-Schwarzschild
backgrounds and a delayed response to shock waves (``switchback effect'')}
\cite{arXiv:2111.02429,arXiv:2210.09647}. Including dependence on matter
fields will not affect those properties.

Consider a complexity functional $\C$ defined for a holographic background
of $D=$10- or 11-dimensional SUGRA 
that asymptotes to AdS$_{d+1}\times X^{D-d-1}$.
In addition to the $D$-dimensional SUGRA, we can also describe the physics of 
the background with a theory defined in the AdS spacetime. This theory 
descends from the SUGRA by truncation or dimensional reduction,
which gives rules relating the $D$-dimensional variables to AdS 
variables.\footnote{The AdS theory is not an effective theory in the Wilsonian
sense unless there is scale separation, which conflicts with some versions
of the AdS distance conjecture \cite{arXiv:1906.05225}.}
Then that translation allows us to rewrite $\C$ as a function of the AdS
variables rather than the SUGRA variables. 

For example, consider the 
$D$-dimensional CV volume complexity, defined as $\C_D = V_{D-1}/G_D\ell$,
where $V_{D-1}$ is the volume of a maximal slice that ends on $\Sigma$, $G_D$
is the $D$-dimensional Newton constant, and $\ell$ is an arbitrary length.
Because the AdS spacetime Newton constant is 
$G_{d+1}=G_D/V_X$ (in terms of the volume $V_X$ of $X$), $\C_D$ is the usual
CV complexity $\C_V=V_d/G_{d+1}\ell$ when the full spacetime is a product
of asymptotically AdS spacetime and $X$. However, $\C_{D}$ is generally 
not the standard AdS spacetime 
CV complexity when the spacetime is not a product.
For example, consider the truncation of $D=11$ SUGRA on $S^7$ 
\cite{hep-th/9903214}; Ref.~\cite{arXiv:2111.14897} showed that $\C_{11}$ is 
the maximal volume in AdS$_4$ weighted by a function of a scalar field
(that describes squashing of the $S^7$ in the full geometry). In fact, 
the integrand in the generalized functional can be exponentially larger
than the volume form. In more general backgrounds, \cite{arXiv:2111.14897}
emphasized that the maximal slices need not be the same for the $D$-dimensional
CV and $(d+1)$-dimensional CV complexities. So a holographic
background embedded in string/M theory has two CV complexities, one of
which may appear to be a distinct generalized complexity in the 
lower-dimensional theory.

On the other hand, we might expect the CA (action) complexity to be the
same in both dimensionalities, i.e., the action remains the action when 
evaluated on the $D$-dimensional background. This should normally be the case,
though it may not, for example, when the Bianchi identities hold only on shell
(as in the $S^7$ truncation of \cite{hep-th/9903214}). An additional question
is whether standard boundary terms in $D$ dimensions always reduce to 
the commonly used boundary terms in AdS spacetime (as described in 
\cite{arxiv:1609.00207}).\footnote{See \cite{arXiv:2405.10363}
for related calculations}
Another issue matching higher- and lower-dimensionality CA complexities
is the origin  of the AdS curvature. If the curvature is (partly) supported
by flux along the AdS directions, as in  
AdS$_3\times S^3\times X^4$ for example, the truncated theory in
the lower-dimensional AdS spacetime 
contains a $(d+1)$-form as well as a cosmological 
constant. While the kinetic term for this top form
is a constant in the action, the coupling to the metric is different than
that for a cosmological constant. Therefore, 
the relationship between the integrand of the action and the AdS scale
is also different. As a result, the AdS spacetime 
CA complexity that matches the 
$D$-dimensional CA functional is not simply the integral of Einstein-Hilbert
and cosmological constant terms but must also include a top form.
(For applications requiring the action evaluated over the full AdS spacetime,
such as calculating the free energy, the variational problem should hold
the top form field strength fixed on the boundary; this requires a boundary
term that equals a compensating bulk term on shell \cite{Duncan:1989ug}.)

\section{Classes of complexity functionals}

We define two classes of complexity functionals, and we
note that there is an analogous classification already for
complexity in AdS backgrounds without extra dimensions.
Recall that there is an arbitrary length scale $\ell$ in the CV complexity
as defined above (as well as in CV2.0 and CA complexities). The common choice
$\ell\sim L$, the asymptotic curvature length (i.e., AdS radius), applies 
equally well to any background with the same AdS asymptotics, as does the
choice $\ell\sim l_P$, the Planck length in the AdS spacetime 
theory. On the other hand,
another common choice for black hole backgrounds is $\ell\sim r_H$, the
horizon radius, which depends on the background --- unless we can write
$\ell$ in a way that interpolates between $L$ or $l_P$ and $r_H$ as a 
function of the background (as suggested in \cite{arXiv:1807.02186}).

Consider a complexity functional $\C_D$ in $D$-dimensional 
SUGRA that applies across different holographic boundary conditions. 
This functional would apply to different AdS curvature scales, different
dimensionality AdS spacetimes, or cascading gauge theories. Such a functional
cannot depend on parameters of any specific background; instead, it should
include only parameters of the $D$-dimensional SUGRA. Likewise, $\C_D$
should have the same invariances as the $D$-dimensional SUGRA.
We will call a functional $\C_D$ that satisfies these criteria 
``universal'' because it should apply to all holographic spacetimes
that are backgrounds within the given $D$-dimensional SUGRA theory.
A universal functional with arbitrary scale $\ell$ must take $\ell\sim l_D$,
the $D$-dimensional Planck length, because that is the only available length
in the SUGRA.\footnote{In 10D, we could choose the string length, but
it is not independent because the ratio of scales is given by a dynamical
variable.} On the other hand, we will call a complexity functional $\C$
that depends on the holographic boundary conditions (such as the AdS scale)
or only obeys the symmetries of the background a ``specific'' complexity
functional.

One question of interest is when an apparently specific functional is actually
universal. Consider again the example of AdS$_4\times S^7$
spacetime. As we argued, the 11D CV functional $C_{11}$ with $\ell=l_{11}$ is
universal, though it appears to be a generalized functional in the 
AdS spacetime theory.
On the other hand, because of a Weyl factor multiplying the AdS components
of the 11D metric, it is not obviously 
possible to write the AdS spacetime CV complexity in
terms of 11D variables in a way that respects 11D diffeomorphism invariance.
Finding universal functionals that reproduce $\C_V$ (or CV2.0, etc)
in a variety of cases is an interesting problem.

We can also ask whether universal functionals can provide new information
about complexity of a given background precisely because they lack knowledge
of the specific solution. For example, consider an AdS$_5\times X^5$ 
background of IIB SUGRA, where $X^5$ has volume $V_X=L^5\Omega_X$.
The universal CV complexity $\C_{10}=V_9/G_{10}l_{10}$
becomes $\C_{10}\sim V_4/G_5L (N/\Omega_X)^{1/4}$ up to factors of 
order unity. This is the usual 5D CV functional rescaled by powers of the
central charge and gauge group rank \cite{hep-th/9807164}. 
In this case, $\C_{10}$ carries 
information about the central charge of the dual gauge theory even in 
backgrounds that are otherwise identical in the AdS dimensions because it
distinguishes the volume of the unit $X^5$.

We do not take a position on the relative value of universal versus 
specific complexity functionals. Indeed, someone who subscribes to the view
that boundary conditions are a fundamental part of the definition of a 
theory of quantum gravity (as in \cite{hep-th/0309170}) may not find a
fundamental difference. However, because of a lack of prior 
literature, our interest here is primarily universal functionals.

\section{Gauge invariance}

As a physical quantity, any complexity functional should be gauge invariant.
We point out here that gauge invariance constrains boundary terms
when a codimension-0 complexity functional has a Chern-Simons (CS) 
term in the integrand (in some cases completely).
While CS terms are ubiquitous in SUGRA actions in many dimensions, we will
focus on action complexity for 10D and 11D SUGRA due to the 
importance of CS terms in that context. For simplicity, we will assume the
absence of any brane sources.

To start, consider the action of 11D SUGRA (the low-energy limit of M theory)
as given by \cite{Cremmer:1978km}
\beq{11sugra}
S_{bulk}=S_{EH}-\frac{1}{16\pi G_{11}}\int_\M\left(\frac 12 F_4\w\star F_4+\frac 16
C_3\w F_4^2\right)\ ,
\eeq
where $F_4=dC_3$ and $S_{EH}$ is the Einstein-Hilbert action.\footnote{There 
is a well-known anomaly term $C_3\w\bar X_8$
\cite{hep-th/9506126}, which is straightforward to include 
in the following because $\bar X_8$ is exact.}
The equation of motion is $d\star F_4=F_4^2/2$.
Under gauge transformations $\delta C_3=d\chi_2$, the CS term yields a 
boundary term in $\delta S$. Normally, we assume $\chi_2$ vanishes on the
boundary of spacetime, so the reader might think we can simply choose $\chi_2$
to vanish on $\del\M$. However, we often want to compare complexities at
different $\Sigma$, meaning $\del\M$ differs.
To make a fair comparison, we cannot change the surface where $\chi_2$ vanishes,
and having $\chi_2$ vanish on all $\del\M$ for all choices of $\Sigma$ means
it will vanish everywhere.
Instead, we take $\chi_2$ to vanish at the spacetime boundary\footnote{Or 
cutoff surface} as usual. 

To make the entire action gauge invariant, we
add the boundary action
\beq{11bdry} S_{bd} =+\frac{1}{48\pi G_{11}}\int_{\del\M} C_3\w\star F_4 \ ,\eeq
where the $\star$ is in the full 11D spacetime and the forms are 
pulled back to $\del\M$. 
This cancels $\delta S$ on shell (where we evaluate complexity) 
because its variation is $\propto \chi_2 d\star F_4=-\chi_2F_4^2/2$. Also on 
shell, we can write $S_{bd}$ as a bulk total derivative, giving the action
\beq{11sugra2} S_{bulk}+S_{bd}=S_{EH}-\frac{1}{16\pi G_{11}}\int_\M\left(\frac 16 
F_4\w\star F_4\right)\ .\eeq

While we must add (\ref{11bdry}) or another boundary term with the 
same gauge variation to the action with the relative coefficient completely 
determined, we may also add any gauge-invariant boundary term.\footnote{However,
if we restrict to the pullback of a bulk 10-form, there are no other 
boundary terms composed solely of $C_3$ and $F_4$.}
Of course, in backgrounds where the CS term vanishes independent of gauge, 
we can ignore this issue and any boundary term,
but those functionals cannot be universal complexities because they are
gauge invariant only for specific backgrounds.

The dimensional reduction of 11D SUGRA on a circle gives the 10D type IIA
SUGRA\footnote{Without Romans mass term, which we leave to future work} 
action with boundary terms, but
considering it as a 10D theory is instructive. One gauge-invariant form for
the bosonic type IIA action is\footnote{For IIA and IIB SUGRAs, we take 
conventions with $d\t F_{p+2} =\t F_{p}H_3$.}
\bea
S &=& S_g+\frac{1}{32\pi G_{10}} \int_\M \left[e^{-2\phi}H_3\w\star H_3 
-F_2\w\star F_2\vphantom{\t F_4}\right.\nonumber\\
&&\left.-\t F_4\w\star\t F_4- C_3\w\t F_4\w H_3\right]\nonumber\\
&&+\frac{1}{32\pi G_{10}} \int_{\del\M}
\left[C_1\w\star F_2+C_3\w\star\t F_4\right]\ .\label{IIASUGRA}\eea
($S_g$ is the string frame dilaton-gravity action.)
However, the boundary term
\bea
&&\int_{\del\M}\left[e^{-2\phi}B_2\w\star H_3+\frac 32 C_1\w\star F_2+\frac 12
C_3\w\star\t F_4\right.\nonumber\\
&&\left. +C_1\w B_2\w\star\t F_4+\frac 12 B_2\w C_3\w dC_3\right]
\label{IIAbdry}\eea
is invariant under gauge transformations of $B_2$ and the Ramond-Ramond 
forms, so any
multiple of it can be added to (\ref{IIASUGRA}). The same is true for the
boundary integral of $\phi\star d\phi$.

The type IIB SUGRA has a slight subtlety related to the self-dual 5-form. 
The usually cited action with kinetic term $\t F_5 \star\t F_5$ is actually
a pseudoaction because that term always vanishes due to the self-duality. 
For an action complexity,
we need an action with only independent components of $\t F_5$. One approach,
when possible, is to identify a coordinate $\t x$ that is orthogonal to all
the others and let $\t f_5$ and $c_4$ be the components of $\t F_5$ and its
potential with no legs along $\t x$. It is often possible to take $\t x$ 
to be one of the coordinates along the conformal boundary. The bulk action is
\bea
S &=& S_g+\frac{1}{32\pi G_{10}} \int_\M \left[e^{-2\phi}H_3\w\star H_3
+F_1\w\star F_1\right.\nonumber\\
&&\left. +\t F_3\w\star\t F_3+\t f_5\w\star\t f_5+c_4\w(\t F_3\w H_3)\right.
\nonumber\\
&&\left. - \t f_5\w(C_2\w H_3)-\t f_5\w\t d c_4\right]\label{IIBSUGRA}\eea
where $\t d=d\t x\del_{\t x}$ (see \cite{arXiv:1907.12755} or fix the 
auxiliary scalar of \cite{arXiv:hep-th/9806140} to find this action). The
boundary action 
\beq{IIBbdry1}
S_{bd} = -\frac{1}{32\pi G_{10}}\int_{\del\M} \left[C_2\w\star\t F_3+c_4\w\t f_5
+c_4\w\star\t f_5\right]\eeq
cancels the boundary gauge variation of (\ref{IIBSUGRA}).
The boundary integrals of $\phi\star d\phi$, $C_0\star F_1$, $\phi\star F_1$, 
and $C_0\star d\phi$, as well as
\bea
S_{bd,2}&=& \int_{\del\M} \left[B_2\wedge\left(e^{-2\phi}\star H_3-C_2\w\left(\t f_5+
\star\t f_5\right)\right.\right.\nonumber \\
&&\left.\left. +C_0\star\t F_3-\frac 12 C_2^2\w H_3\right)-C_2\w\star\t F_3
\right]\label{IIBbdry2}
\eea
are gauge-invariant boundary terms, so it is possible to include them with
arbitrary coefficient.

We note that some SUGRA theories in lower dimensions also have CS terms, so
similar considerations apply even in the AdS spacetime 
theory. The usual CA complexity
in backgrounds with nonvanishing CS term is not gauge invariant unless
supplemented with appropriate boundary terms.

We emphasize that the boundary terms discussed here are not the same boundary 
terms that change the variational problem from having fixed potential at
the boundary to fixed field strength. When numerical prefactors are given,
they differ for the two applications.

\section{Proposed complexity functionals}
Here we propose new universal complexity functions motivated by the discussion
above.

As noted, it is common to set the arbitrary scale of CV or CV2.0 complexity 
in the AdS spacetime 
theory to the AdS length, $\ell=L$; a nice feature of this choice
is that CV complexity obeys (a modified version of) Lloyd's bound
\cite{Lloyd:2000cry,arXiv:2109.06883}. This choice lifts to a specific
complexity functional for product spacetimes AdS$_{d+1}\times X^{D-d+1}$.
It turns out that there also exist universal
complexity functionals, which we call CF and CF2.0, that reproduce CV and CV2.0
with $\ell=L$ in many cases for AdS$_{d+1}\times X^{D-d+1}$ products.
The key observation is that 
these spacetimes are supported by one or more field strengths threading the 
AdS or $X$ directions (or both) with square $|F|^2\propto 1/L^2$. The 
complexity functionals are
\beq{cf1}
\C_F \equiv \frac{1}{G_D}\int_{\mathcal{B}} d^{D-1}x\sqrt{|g|}\sqrt{\mathcal{F}}
\ ,\
\C_{F2} \equiv \frac{1}{G_D}\int_{\M} d^{D}x\sqrt{|g|}\,\mathcal{F}\ ,\eeq
where $\mathcal{B}$ is the codimension-1 surface that optimizes the integral,
and $\M$ is the Wheeler-DeWitt patch (the same integration region as CV2.0).
The integrands are given by $\mathcal{F}_M\equiv a_M\| F_4\|^2$ in M theory and
\beq{cf2} \mathcal{F}_{IIB} \equiv e^{q\phi}\frac{a_1}{l_{10}^2}+a_2\left(e^{2\phi}
\|\t F_3\|^2 +\|H_3\|^2\right) +a_3 e^{2\phi}\|f_5\|^2 
\eeq
in the IIB SUGRA string 
frame\footnote{$\|F_p\|^2\equiv |F_{\mu_1\cdots\mu_p}F^{\mu_1\cdots\mu_p}|/p!$}
for arbitrary constants $a_i$ and $q>1/2$. The first term in $\mathcal{F}_{IIB}$ 
is the only contribution when the contracted field strengths vanish, 
as in the AdS$_3\times S^3\times X^4$ backgrounds \cite{hep-th/9804085},
but it is very small for $q>1/2$ ($\sim g_s^{q-1/2}(g_s\sqrt{Q_1Q_5})/L^2$, 
where $g_s\sqrt{Q_1Q_5}$ is large and fixed with $g_s\to 0$).
In AdS$_{d+1}\times X^{D-d+1}$ spacetimes, these ``flux complexities'' 
reproduce the volume complexities for geometric backgrounds; for example,
$a_1=0,a_3=1/16$ gives the usual volume complexities for AdS$_5\times S^5$.
However, CF and CF2.0 complexities generally differ from CV and CV2.0
in charged backgrounds when the electromagnetic field descends from a 
$D$-dimensional form. Similarly, two charged black holes that appear identical
in AdS spacetime may have 
different CF and CF2.0 complexity when the corresponding electromagnetic
fields have different $D$-dimensional origins (in contrast to a complexity
functional of the AdS spacetime theory).

The considerations of the above section also motivate us to define a new 
version of action complexity, which we dub CA2.0 and can apply in any theory
with CS terms. Like the usual CA complexity, 
CA2.0 evaluates the action over the Wheeler-DeWitt patch, 
taking the usual gravitational boundary terms \cite{arxiv:1609.00207}
as well as boundary terms for the forms
that make the entire functional gauge invariant on shell. There may also be
freedom to add a gauge-invariant boundary term with arbitrary coefficient,
and we suggest a particular value for the IIA and IIB SUGRAs here. Recalling
that the on-shell form boundary terms can be written as bulk terms, we 
propose functionals
\beq{ca2M}
\C_{A2} \equiv S_{grav}-\frac{1}{16\pi G_{11}}\int_\M\left(\frac 16 
F_4\w\star F_4\right) \eeq
for M theory and
\bea
\C_{A2} &\equiv& S_{grav}+\frac{1}{32\pi G_{10}} \int_\M \left[\frac 13
e^{-2\phi}H_3\w\star H_3-F_2\w\star F_2\right.\nonumber\\
&&\left.-\frac 13\t F_4\w\star\t F_4\right]\label{ca2IIA}
\eea
for type IIA SUGRA, which includes a multiple of \eqref{IIAbdry} chosen to 
give the dimensional reduction of \eqref{ca2M}. Here, $S_{grav}$ is either the
Einstein-Hilbert or dilaton-gravity bulk action supplemented with the usual
gravitational boundary terms. In type IIB SUGRA, we choose boundary terms 
consistent with S-duality, giving\footnote{Interestingly, the 5-form does not 
contribute, so the complications due to its self-duality have disappeared.}
\bea
\C_{A2} &=& S_{grav}+\frac{1}{32\pi G_{10}} \int_\M \left[F_1\w\star F_1
+\frac 12 e^{-2\phi}H_3\w\star H_3\right.\nonumber\\
&&\left. +\frac 12 \t F_3\w\star\t F_3\right]\ .\label{ca2IIB}
\eea
While the choice of boundary terms in complexity for form fields may not always
be clear, for theories with Chern-Simons terms, a universal action complexity
must include boundary terms that preserve gauge invariance.

One other comment on action complexity is important. For 
product AdS$_{d+1}\times X^{D-d+1}$ backgrounds, it is the usual SUGRA 
action in $D$ dimensions without boundary terms for $B_2$, $C_{p+1}$ 
that reduces to the CA complexity on AdS$_{d+1}$ (with action given by 
Einstein-Hilbert, cosmological constant, and boundary terms).
For example, with flux only on $X$, the Ricci scalar of $X$ plus the flux 
kinetic term give exactly the $d(d-1)/L^2$ cosmological constant term for
the gravity action in AdS spacetime. 
The CA2.0 functional is clearly distinct, and the usual CA complexity
may not be gauge invariant for some backgrounds when considered in the 
context of the higher-dimensional gauge/gravity duality.

\section{Discussion}

We have discussed several issues related to holographic complexity in the 
full 10- or 11-dimensional gauge/gravity duality: the formulation of 
higher-dimensional complexity functionals in the lower-dimensional 
AdS spacetime theory,
conditions when a complexity functional can apply universally to holographic
backgrounds as opposed to specific asymptotic conditions, and gauge
invariance for Chern-Simons terms. In light of these considerations, we have
also proposed several new universal complexity functionals. All of these
topics require further development for a theory of complexity in string and
M theory. The behavior of the CF, CF2.0, and CA2.0 functionals in 
previously studied backgrounds, such as a variety of AdS black holes, is
of interest; checking explicitly that these functionals satisfy
linear late-time growth for black holes and the switchback effect is an
important point left to future work.
It is also important to consider complexity in
general gauge/gravity dualities beyond the AdS/CFT correspondence 
(see \cite{arXiv:2311.18804} for first steps). 

There are other outstanding questions. For example, comparing complexity of
different backgrounds (e.g. complexity of formation, which subtracts the
complexity of empty AdS spacetime) requires matching cutoff surfaces at the same
Fefferman-Graham coordinate. However, in the full theory, the metric dependence
on the holographic radial coordinate can vary with orientation in the
additional $X^{D-d-1}$ space. The proper definition of the cutoff surface in
such an anisotropic case has yet to be determined and will appear in future
work.

Assuming de Sitter compactifications of string theory exist, the
considerations discussed here will also apply to holographic complexity in 
those de Sitter backgrounds (following 
\cite{arXiv:2202.10684,arXiv:2304.15008}). Indeed, given the many ingredients
needed in proposed de Sitter backgrounds of string theory, there are likely
further subtleties for holographic complexity in cosmology.

Finally, we note that 
some of the observations here may apply to other information
theoretic quantities when considered as part of the full gauge/gravity duality
as opposed to the lower-dimensional AdS spacetime only.

\begin{acknowledgments}
A.R.F. would like to thank N.~Agarwal, M.~Grehan, F.~Seefeld, and P.~Singh 
for discussions
while collaborating on related work. This work has been supported by the
Natural Sciences and Engineering Research Council of Canada Discovery Grant 
program, Grant No.~2020-00054.
\end{acknowledgments}

\bibliography{complexityfull}

\end{document}